# Spontaneous Emission from Electronic Metastable Resonance States

Amir Sivan[*1,2], Milan Šindelka[†3,4], Meir Orenstein[1,2] and Nimrod Moiseyev[2,4,5]
[1]Andrew and Erna Viterbi department of Electrical & Computer Engineering, Technion – Israel Institute of Technology, Haifa 32000, Israel
[2]Helen Diller Quantum Center, Technion – Israel Institute of Technology, Haifa 32000, Israel
[3]Department of Chemistry, Guangdong Technion Israel Institute of Technology, Shantou, Guangdong Province 515603, P.R. China
[4]Schulich Faculty of Chemistry, Technion – Israel Institute of Technology, Haifa 32000, Israel
[5]Faculty of Physics, Technion – Israel Institute of Technology, Haifa 32000, Israel

**Abstract:** We demonstrate that calculating the spontaneous emission decay rate from metastable resonance states (states with finite lifetimes embedded in the continuum) requires considering transitions to all continuum states, not just to lower states. This holds even when the lifetimes of the metastable states are very long and might be effectively considered as bound states in the continuum. However, employing complex-scaling transformations, this computationally prohibitive task becomes feasible by utilizing methods originally designed for excited bound states for calculation of complex poles of the scattering matrix. As an illustrative example, these methods are applied to calculate the spontaneous emission decay rates of metastable resonance states in a double-barrier potential. The rapid numerical convergence of this approach highlights a new avenue for studying spontaneous emission from metastable states in real-life systems, particularly in many-electron systems, where calculation of the spontaneous emission decay rate from metastable resonances (e.g., autoionization states) is computationally difficult, if not impossible, using the standard (Hermitian) formalism of quantum mechanics.

The existence of bound states in the continuum (BSCs) has been known and extensively studied since proposed by von Neumann and Wigner in the early times of quantum mechanics [1]. These infinitely long-lived BSCs become metastable resonance states with finite lifetimes for any infinitely small perturbation that couples the internal degrees of freedom of the system under study and breaks the separability of the original Hamiltonian (see for example Ref. [2] and references therein). These metastable resonance states are associated with the isolated complex poles of the scattering matrix which are sufficiently embedded near the real axis in the complex energy plane as described, for example, in the seminal book of Taylor on scattering theory [3]. Resonance phenomena are related to a myriad of processes – resonant tunneling [4-6], cold-atom scattering [7], photon-phonon scattering [8], the Stark effect [9], high-harmonic generation [10-12] and generally any state undergoing nonradiative processes. For more resonance phenomena in nature see Chapter 2 in [13].

Spontaneous emission (SE) is a fundamental phenomenon central to the fields of quantum optics and quantum field theory [14,15]. It occurs when a quantum matter (atom, molecule, quantum dot or well, etc.) is coupled to the electromagnetic vacuum and spontaneously a emits photon while transitioning from an excited state to a lower energy state. While optical transitions were studied thoroughly in the framework of conventional Hermitian quantum mechanics [16], yielding the Fermi golden rule for transitions between electronic bound states, between bound states and continuum and between continua [17], they do not inherently accommodate many realistic situations where the excited states are metastable states which have finite lifetimes due to ionization, tunneling, coupling to phonons and a variety of other nonradiative mechanisms. This is because the description of open systems is not native to the conventional Hermitian formulation of quantum-mechanics. Such cases are usually treated by applying quantum dissipation theory to the electronic decay channels, e.g. by Lindblad operators governing the electronic master equation or by involved dressed states that are introduced to the problem as a macroscopic reservoir "absorbing" the electron energy [18-23]. These methods are usually complicated both theoretically and computationally, often making the interplay between SE and various physical decay mechanisms obscure.

Here, we comprehensively address the question of the effect of the resonance widths (i.e., inverse lifetimes) characterizing resonant states on the phenomenon of SE. Using the non-Hermitian quantum-mechanics (NHQM) perspective that is well-suited to describe metastable resonance states [13], we present closed form expression for the SE decay rate. This formalism treats the decay channels of the electron dissociation and the photon emission on equal footing [24], and enables the systematic construction of the corresponding resonance states (wavefunctions, energies, lifetimes). We remark that very few works have applied the NHQM formalism to problems pertaining to SE (see Ref. [24]).

Our letter is organized as follows: First, we introduce the expression we have obtained for the SE decay rates as a function of the native lifetime of the electronic metastable state, emphasizing the need to include all transitions from the initial resonance state to all bound and continuum states. This contrasts with only considering transitions to lower bound states and the part of the continuum with lower energy than the initial state, as in the case normally considered for SE where the initial state is approximated as a bound state.

We then emphasize the dependence of this expression on our choice of the complex-scaling (CS) transformation angle $\theta$ by which the electronic coordinates are rotated into the lower half of the complex plane, as part of the CS method we use in order to treat both the non-Hermitian electron resonances [13,25] and the photon emission processes [24]. In our framework, we address the two distinct decay mechanisms by complex-scaling the electronic and photonic degrees of freedom by different angles. Further details on the CS method are provided in the Supplementary Material [26]. We choose this method because, in our problem, non-Hermiticity arises from both SE due to coupling with the electromagnetic bath, *and* the outgoing boundary conditions of the electronic wavefunctions – a loss mechanism that is unrelated to coupling to a reservoir. CS treats both decay processes in a unified manner, unlike, for example, the Lindblad formalism, which is inadequate for describing loss that does not arise from coupling to an external environment. In the Hermitian Hamiltonian picture, metastable states are associated with a locally high density of states in the continuum [27]; as the transformation angle $\theta$ is increased, additional complex poles of the scattering matrix, describing these metastable states and their density of states in the continuum, emerge from the rotated continuum. These poles correspond to square-integrable wavefunctions embedded in the generalized Hilbert space. Here, $\theta = 0$ implies the use of the standard Hermitian quantum-mechanical formalism, where the electronic spectrum consists only of bound and continuum states. Importantly, our result is reduced to the known expression obtained by the conventional Fermi golden rule for SE from a bound excited state to lower-energy bound states of the atomic, molecular or other system under study.

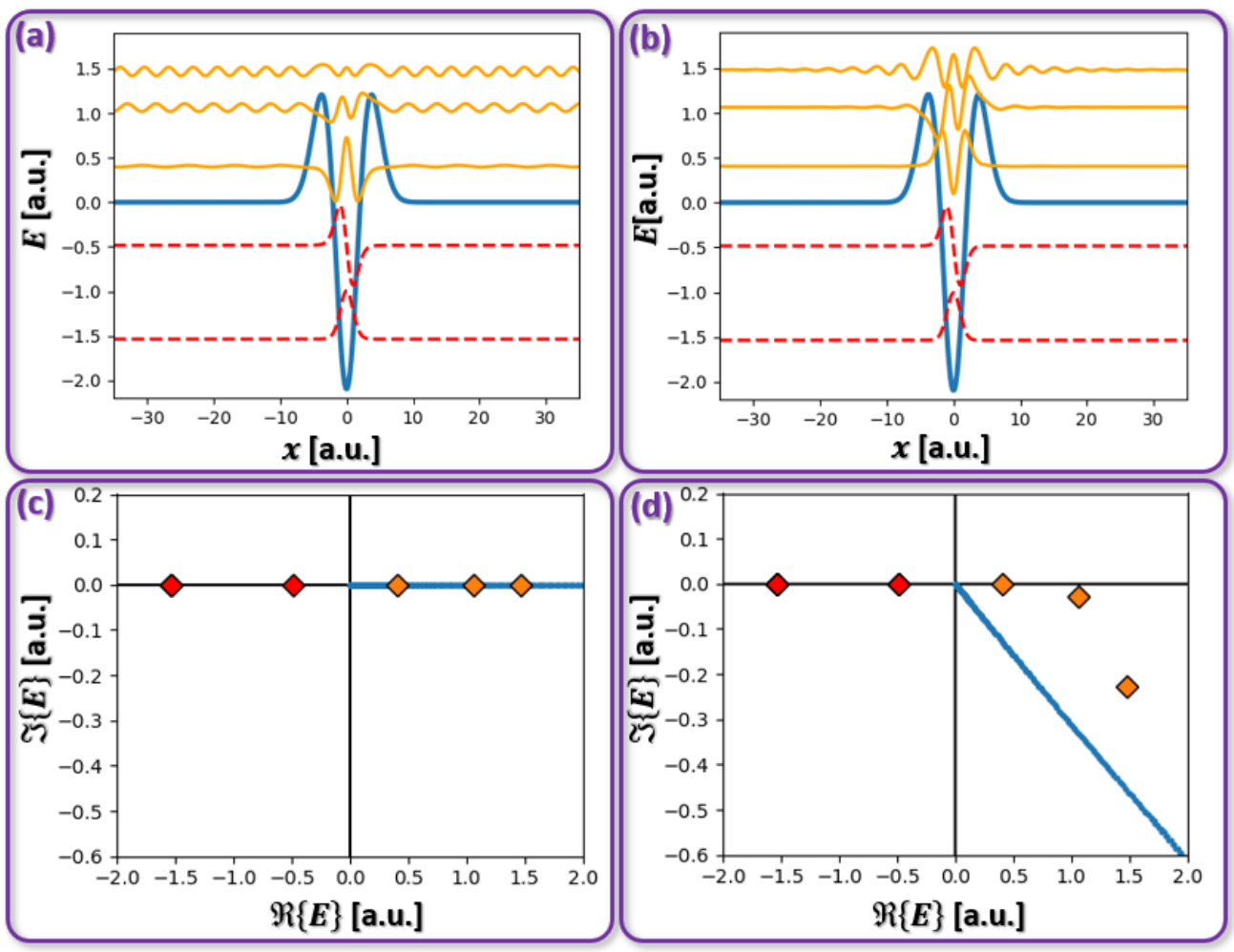

Fig. 1. Electronic wavefunctions and complex spectra of the electronic potential $V(x) = (0.5x^2 - 2.1)e^{-0.1x^2}$ (in a.u. – atomic units). (a,b) Real parts of the wavefunctions of the two bound states (red) and the longest-lived three resonance states (orange), supported by the (unscaled) potential (blue). In (a) the Hamiltonian is solved for $\theta = 0$ (Hermitian case) and the resonance states are unbounded as evidenced by their oscillations, whereas in (b) complex-scaling is applied with a rotation angle $\theta = 0.15[rad]$ and the resonance states become square-integrable as seen by their decaying envelopes. (c,d) Complex spectrum of the electronic states for the Hermitian and $\theta = 0.15[rad]$ cases, respectively. Red (orange) diamonds denote the discrete bound (resonance) states and blue line marks the continuum.

Lastly, we demonstrate the rapid convergence of the calculated SE decay rate from a metastable resonance state as the number of complex resonance poles, exposed by increasing $\theta$, is increased. For elucidating the concept, we will use a one-dimensional double-barrier potential as an illustrative example, as shown in Fig. 1. However, the conclusions hold for any potential in any multidimensional many-particle setup.

In our model the field is fully quantized, and we calculate the SE decay rate by applying 2nd-order perturbation theory on a second-quantized momentum-gauge Hamiltonian (see Supplementary Material [26]). By employing CS and decomposing the continuum, we find that the SE decay rates from an electron resonance state consists of contributions from the entire rotated continuum and all discrete states (bound and resonance), including those with higher energies than the initial state energy. The total SE rate from the discrete state $n$ is given by

$$\gamma_n = \sum_{n'}^{N(\theta)} \Delta\gamma_{nn'} + \left(\int_{l_{2\theta}} d\mathcal{E}\Delta\gamma_n(\mathcal{E})\right)\left(N(\theta)\right) \quad (1)$$

where the first term consists of the contributions $\Delta\gamma_{nn'}$ to the SE partial decay rates from the state $n$ to the discrete states $n'$. Note that the summation is over the indices $n' \in \{1,\dots,N(\theta)\}$ with the total number of discrete states $N(\theta)$ depending on the CS angle $\theta$. These discrete states consist of bound states, and resonance states associated with the emergent complex poles which we utilize as a partial basis set for the continuum. It may seem less intuitive that we include terms of emission from higher discrete energy states than the real energy of the initially excited resonance. However, we should recall that excitation of a resonance state represents excitation to an unbound (Lorentzian) continuum distribution represented by its discrete complex pole. Thus, the initial excitation of a resonance state includes partial excitation of all energy states with real energy larger than the real energy of the nominally excited resonance and their emission contributes to the total spontaneous emission decay rate.

The second term of (1) describes the contribution of the rotated continuum, where $d\mathcal{E}\Delta\gamma_n(\mathcal{E})$ denotes the infinitesimal contribution around the complex energy $\mathcal{E}$ along the positive real semi-axis rotated clockwise by $2\theta$ denoted by the line $l_{2\theta}$. This of course contains an infinitely long part of the continuum that is more energetic than the emitting resonance. Since the total decay rate $\gamma_n$ is an observable and thus independent of $\theta$, the result of the integral must change whenever another complex pole emerges from the continuum and therefore the second term depends on $\theta$ through $N(\theta)$.

The partial contributions from the transitions to other discrete states are given by

$$\Delta\gamma_{nn'} \equiv -\frac{1}{2\pi c}\left[\operatorname{Im}\left(\mathrm{Z}_{nn'}\right)\log\frac{\left|f_{nn'}\right|}{c^2} + \operatorname{Re}\left(\mathrm{Z}_{nn'}\right)\arg\left(-f_{nn'}\right)\right] \quad (2)$$

where $\mathrm{Z}_{nn'} = d_{nn'}^2 f_{nn'}$ and we have denoted the complex energy difference $f_{nn'} = \omega_{nn'} - i\Gamma_{nn'}/2$. The differences in frequency and decay rates are respectively denoted in the above by $\omega_{nn'} = \omega_n - \omega_{n'}$ and $\Gamma_{nn'} = \Gamma_n - \Gamma_{n'}$, and $d_{nn'}$ is the transition dipole moment matrix element (which is a complex number for metastable states). The infinitesimal contributions comprising the integral term of (1) are similarly given by

$$d\mathcal{E}\Delta\gamma_n\left(\mathcal{E}\right) \equiv -\frac{d\mathcal{E}}{2\pi c}\left[\operatorname{Im}\left(\mathrm{Z}_n\left(\mathcal{E}\right)\right)\log\frac{\left|f_n\left(\mathcal{E}\right)\right|}{c^2} + \operatorname{Re}\left(\mathrm{Z}_n\left(\mathcal{E}\right)\right)\arg\left(-f_n\left(\mathcal{E}\right)\right)\right] \quad (3)$$

with the corresponding continuous functions defined by $\mathrm{Z}_n\left(\mathcal{E}\right) = d_n^2\left(\mathcal{E}\right) f_n\left(\mathcal{E}\right)$, and the complex energy difference for the case of transition into a continuum state is $f_n\left(\mathcal{E}\right) = \tilde{\omega}_n\left(\mathcal{E}\right) - i\tilde{\Gamma}_n\left(\mathcal{E}\right)/2$ with the complementary definitions of frequencies and decay rates differences, namely $\tilde{\omega}_n\left(\mathcal{E}\right) = \omega_n - \operatorname{Re}\left(\mathcal{E}\right)$ and $\tilde{\Gamma}_n\left(\mathcal{E}\right) = \Gamma_n + 2\operatorname{Im}\left(\mathcal{E}\right)$. $d_n\left(\mathcal{E}\right)$ is the transition dipole moment between a discrete state $n$ and a continuum state with energy $\mathcal{E}$. The derivation of (1)-(3) is given in [26]. Expressions (2) and (3) establish the relations between the SE rate and the finite lifetimes of the resonance states due to nonradiative mechanisms.

For $\theta = 0$ (the Hermitian case) the equations given above are reduced to the well-known Fermi's golden rule for a spontaneous emission from an excited bound state. In this case $f_{nn'} = \omega_{nn'}$ and the dipole transition moments are real quantities. Then, under a proper choice of branch, $\arg\left(-f_{nn'}\right)$ equals $\left(-\pi\right)$ if $n > n'$ and zero otherwise and $\operatorname{Re}\left(\mathrm{Z}_{nn'}\right) = \omega_{nn'}\left|d_{nn'}\right|^2$, so eq. (2) reverts to the known result obtained for SE in the Hermitian formalism in 1D,

$$\Delta\gamma_{nn'}^0 = \frac{\omega_{nn'}}{2c}\left|d_{nn'}\right|^2. \quad (4)$$

Since there is no continuum below the threshold energy, the continuum does not contribute, so the total SE rate from a bound state is simply the sum of its decay rates to lower levels. We remark that the energy shifts (Lamb shifts) associated with the coupling to the electromagnetic vacuum are still affected by higher-energy bound states and the continuum as predicted by the conventional Hermitian quantum-mechanical approach [28] (see Supplementary Material [26]).

The main result of this letter is the rapid convergence of the calculation of the SE decay rate from a metastable resonance state that is obtained by considering the contributions of only few neighboring complex discrete states that are separated from the continuum by the judicious selection of the angle $\theta$. The profound significance of our main finding is evidenced from equation (1), where we see that every contribution of a discrete state (continuum) requires the calculation of $Z_{nn'} = d_{nn'}^2 f_{nn'}$ $\left(\mathrm{Z}_n\left(\mathcal{E}\right) = d_n^2\left(\mathcal{E}\right) f_n\left(\mathcal{E}\right)\right)$. The calculations of the transition dipole moments $d_{nn'}$ and $d_n\left(\mathcal{E}\right)$ require obtaining the complex-scaled electronic wavefunctions of the discrete states and the continuum, respectively, and integrations over the entire coordinate space (which almost always must be performed numerically). Therefore, while the calculation of several discrete states contributions is simple, calculating the contribution of a "continuum" approximated by a very dense spectrum of states is computationally exorbitant. These calculations become particularly prohibitive when considering multi-dimensional coordinate spaces describing realistic many-electron 3D systems.

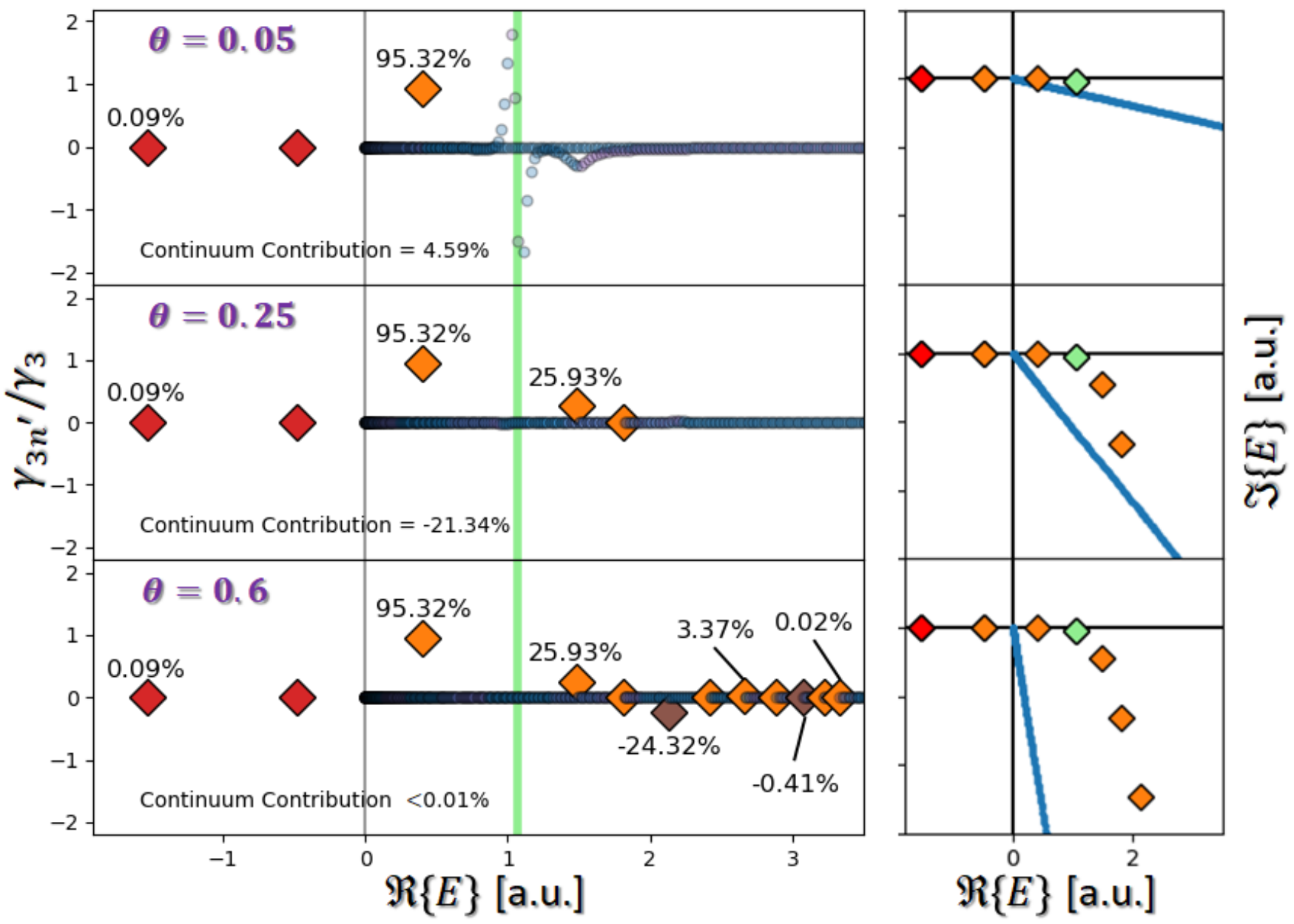

Fig. 2. Left panel: decomposition of the total SE decay rate $\gamma_3$ of the resonance state $n = 3$ into partial decay rates $\Delta\gamma_{3n'}$ and $\delta\gamma_3(E)$ to the discrete states and the rotated continuum, respectively, for different rotation angles $\theta$. Green vertical line denotes the real energy of the resonance state $n = 3$. Red diamonds denote bound states, orange (brown) diamonds denote electronic resonance states with positive (negative) contribution to the total decay rate $\gamma_3$. It is shown that over 99.99% of the SE decay rate can be calculated from only the first twelve resonances. Right panel: the complex energy spectra for the rotation angles $\theta$ corresponding to the left panel.

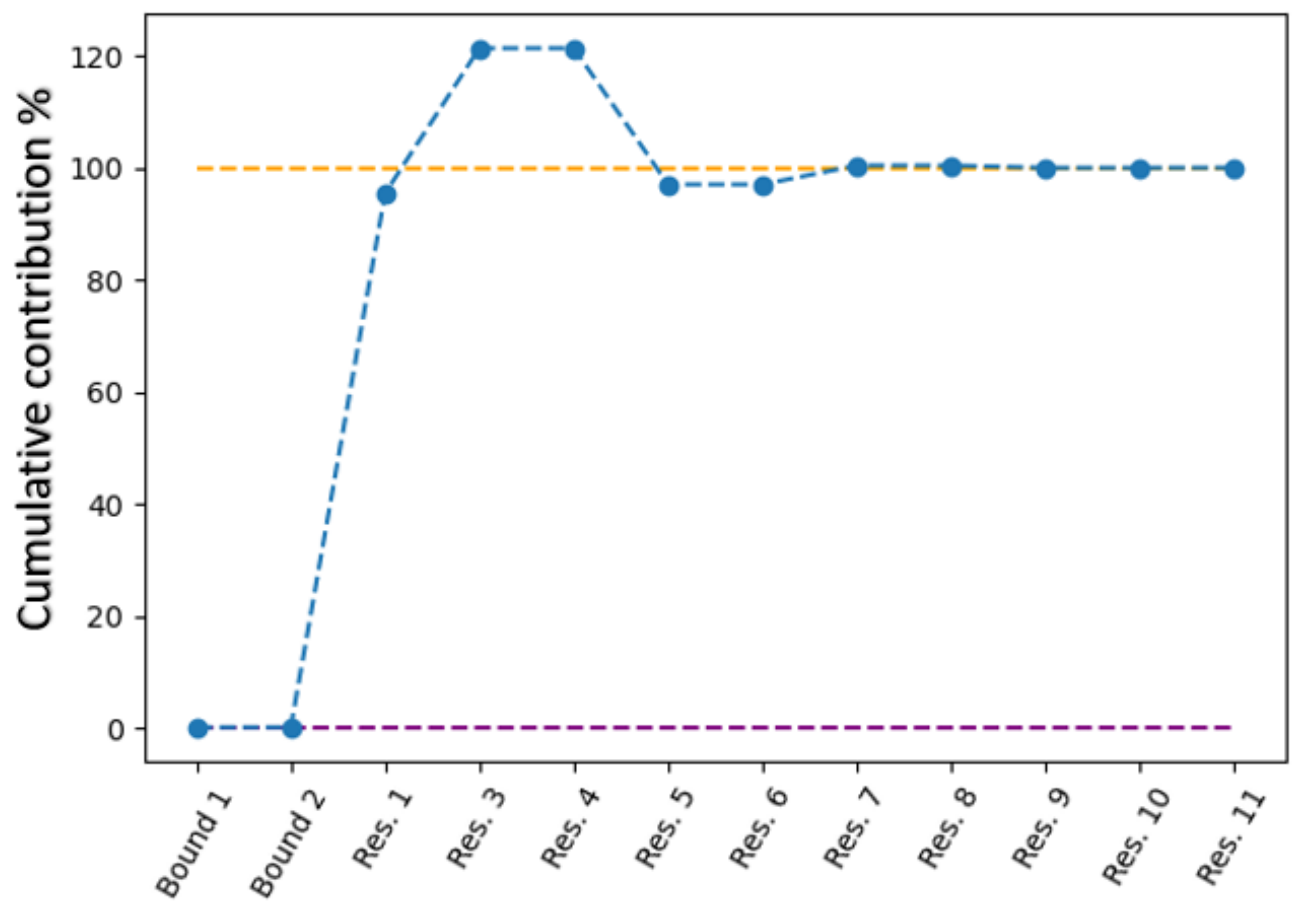

Fig. 3. Cumulative contributions of the discrete states to the total decay rate of the second resonance, $\gamma_3$. Shown here is the rapid oscillatory convergence of the calculation of $\gamma_3$ to its exact value from the first twelve discrete states appearing at the bottom panel of Fig. 2.

To demonstrate this main finding, we examine the SE process from the second resonance state of an electron system defined by the potential $V(x) = (0.5x^2 - 2.1)e^{-0.1x^2}$ illustrated in Fig. 2. A similar potential was used to describe resonant tunnelling in a diode [29]. We show how each electronic resonance state that emerges from the continuum as $\theta$ is increased contributes to the total SE decay rate $\gamma_3$, at the expense of the contribution of the remaining electronic continuum. Expressly, the discrete contributions $\sum_{n'\neq 3}^{N(\theta)} \Delta\gamma_{3n'}$ from the first eight (ten, twelve) discrete states account for approximately 99.57% (99.975%, 99.996%) of the total decay rate $\gamma_3$. Note that the odd discrete states do not contribute to the total rate $\gamma_3$ due to the spatial symmetry of our potential function $V(x)$. Our method therefore produces excellent approximations requiring only a few relatively straightforward and inexpensive calculations. Notably, the contributions of the discrete poles of the scattering matrix on the SE decay rate become smaller farther away from the emitting resonance state (see Fig. 3); this is expected as the spectral shapes of distant resonance states have smaller overlaps and their respective transition dipole moments become smaller. An example of a possible realistic implementation of spontaneous emission from a similar potential describing a semiconductor based finite quantum well with finite tunnelling barriers is discussed in [26].

In conclusion, we have extended the calculations of spontaneous emission decay rates from bound states to metastable finite-lifetime states (resonances with finite lifetimes embedded in the continuum). Our comprehensive model predicts nontrivial dependence of the SE decay rate on the nonradiative lifetime of the electron. The finite lifetime of excited states can arise from various mechanisms, such as autoionization, predissociation, or interaction with the environment (e.g., phonon coupling) – all of which are prevalent in realistic physical systems. Calculating the spontaneous emission decay rate from metastable resonance states is practically impossible in most of these realistic problems using the standard Hermitian quantum mechanics formalism, as these states are, according to the conventional formalism, distributed over the entire continuum and therefore calculating the pertinent SE rates requires considering transitions to all continuum states and not just to lower bound states. However, as we demonstrated here, by summing over the complex poles of the scattering matrix using complex-scaling transformations, this computationally prohibitive task becomes feasible. We remark that the single electron systems on which we focused in this paper to elucidate our model do not pose a limitation on its generality, as single-electron models are also used to represent many-electron phenomena, e.g. high-harmonic generation [12]. Furthermore, many-electron effects (e.g., Auger effect) can in principle also be incorporated into the model at the cost of higher numerical complexity. Thus, we provide a new avenue for the treatment of spontaneous emission processes that occur in open systems, which is highly relevant for treatment of all realistic light sources – from quantum dots to lasers.

A. S. acknowledges partial fellowship from the Helen Diller Quantum Center of the Technion – Israel institute of Technology. M. Š. acknowledges generous financial support of the Helen Diller Quantum Center of the Technion – Israel institute of Technology (Dec 21, 2023 – Sep 17, 2024). N. M. acknowledges the Israel Science Foundation (ISF) Grant No. 1757/24 for partial support.

Corresponding authors:
*amirsi@campus.technion.ac.il
†milan.sindelka@gtiit.edu.cn

# Spontaneous Emission from Electronic Metastable Resonance States

## Supplementary Material

Amir Sivan*[1,2], Milan Šindelka†[3,4], Meir Orenstein[1,2] and Nimrod Moiseyev[2,4,5]
[1]Andrew and Erna Viterbi department of Electrical & Computer Engineering, Technion – Israel Institute of Technology, Haifa 32000, Israel
[2]Helen Diller Quantum Center, Technion – Israel Institute of Technology, Haifa 32000, Israel
[3]Department of Chemistry, Guangdong Technion Israel Institute of Technology, Shantou, Guangdong Province 515603, P.R. China
[4]Schulich Faculty of Chemistry, Technion – Israel Institute of Technology, Haifa 32000, Israel
[5]Faculty of Physics, Technion – Israel Institute of Technology, Haifa 32000, Israel

### A. Complex-scaling of atomic and photonic coordinates

Here we outline the complex-scaling (CS) technique of the non-Hermitian quantum-mechanics (NHQM) formalism in the context of our work. The technique is detailed thoroughly in chapter 5 of Ref. [1], and in Ref. [2] for application to photonic coordinates, and here we will only mention important results relevant to our work without proving them.

We scale the electronic or photonic coordinates by a complex quantity $\eta = e^{i\theta}$ in order to ensure that the solutions with outgoing boundary conditions are square-integrable. A coordinate $x$ will be transformed as $x \mapsto x\eta = xe^{i\theta}$, where the parameter $\theta \in \mathbb{R}$ is an angle measured clockwise. We emphasize that this parameter is a mathematical tool we employ in order to treat divergent wavefunctions, and is not a physical observable. Additionally, we remark that CS effectively introduces a reflectionless complex absorbing potential [1] via the transformation $V(x) \to V(xe^{i\theta})$.

To treat the two distinct decay channels described in our problem – namely, electron dissociation and spontaneous emission (SE), we complex-scale both electronic and photonic coordinates. This amounts to defining scaling operators for the electronic coordinates $\hat{S}_{A,\theta}$ by an angle $\theta$ such that $\hat{S}_{A,\theta}\psi(x) = e^{i\theta/2}\psi(xe^{i\theta})$, and scaling operators for the photonic coordinates $\hat{S}_{R,\phi}$ by an angle $\phi$ such that $\hat{S}_{R,\phi}\hat{a}(k)S_{R,\phi}^{-1} = e^{-i\phi/2}\hat{a}(ke^{-i\phi})$, where $\psi(x)$ is the electronic wavefunction and $\hat{a}(k)$ is the annihilation operator of a photon with a wavenumber $k$ [1,2]. The energy spectrum of the rotated Hamiltonian $\hat{H}_{\theta,\phi} \equiv \hat{S}_{A,\theta}\hat{S}_{R,\phi}\hat{H}\hat{S}_{R,\phi}^{-1}\hat{S}_{A,\theta}^{-1}$ has a complex energy spectrum consisting of discrete states with both real- and complex-valued energies and complex-valued rotated electronic and photonic continua. In order for the continua to be equally rotated, since the electronic continuum rotates as $2\theta$ and the photonic continuum as $\phi$ (see for example [1,2]), we impose $\phi = 2\theta$ and denote the rotated Hamiltonian by $\hat{H}_\theta \equiv \hat{H}_{\theta,2\theta}$. This choice of equally-rotated continua becomes meaningful at later stages of the derivation of SE when contour integration along the continua is necessary; a choice of non-parallel continua will introduce a 2D continuum rather than a 1D continuum, which is mathematically more complicated. An illustration of the entire spectrum (electronic and photonic) containing both continua is given in Fig. S1 in Supplementary Section C for the Hamiltonian under study in this work.

### B. Spontaneous emission from electronic bound states in the NHQM formalism

To lay down the mathematical foundations required for the analysis of SE from electronic resonance states (ERS), we start by considering SE from electronic bound states (EBS) by using the NHQM formalism of CS [1,2]. Our goal is to obtain the known results for the SE decay rate and energy shift by using the CS method, which we will then generalize to the ERS, in order to familiarize the reader with the CS method and our notations.

For simplicity, we consider here a 1D problem. However, the results obtained are easily generalizable to any number of dimensions and coordinates, required to describe any multielectron system of interest.

We consider the standard momentum gauge Hamiltonian

$$\hat{H} = \frac{1}{2m}\left(\hat{p} - \frac{q}{c}\hat{A}(x)\right)^2 + V(x) + \hat{H}_R \tag{S1}$$

with the 1D vector potential [2]

$$\hat{A}(x) = \int_{-\infty}^{\infty} dk \sqrt{\frac{\hbar c^2}{8\pi\omega_k}}\hat{a}(k)e^{ikx} + cc. \tag{S2}$$

and the radiation Hamiltonian

$$\hat{H}_R = \hbar \int_{-\infty}^{\infty} dk\omega_k \hat{a}^\dagger(k)\hat{a}(k). \tag{S3}$$

Henceforth, we will assume the dipole approximation, namely – we will neglect the spatial distribution of the vector potential operator (S2) such that $\hat{A} \equiv \hat{A}(0)$. In the above, the operator $\hat{p}$ is the momentum operator (referring to the momentum in the $x$ direction in our 1D problem), $m$ is the electronic mass, $q$ is the electronic charge, $\hbar$ is the reduced Planck's constant and $c$ is the speed of light in vacuum. The operator $\hat{a}(k)$ denotes the usual operator annihilating a single photon in the mode $k$, obeying the commutation relations

$$[\hat{a}(k), \hat{a}^\dagger(k')] = \delta(k - k'). \tag{S4}$$

We emphasize that in the NHQM formalism, dissociation of a particle from the system (whether it is an electron or a photon) is associated with a complex-valued energy and its corresponding eigenfunction. Therefore, the electronic resonance and the photon emission are treated in the same way – by imposing outgoing boundary conditions on the wavefunctions of the dissociating particles (the concept of photonic coordinates and the photonic wavefunction is discussed e.g. in Refs [3,4]). The NHQM formalism allows one to avoid the need to solve the time-dependent Schrödinger's equation as in the conventional formalism of (Hermitian) quantum-mechanics and enables us to obtain cross-sections and transitions probabilities by solving the time-independent Schrödinger's equation for the non-Hermitian Hamiltonian. See chapter 8 of Ref. [1] and references therein, e.g. [5]. Therefore, in this paper we treat the SE process by using NHQM which enables us the use 2nd-order perturbation theory (2PT) (although we may formulate an exact solution [2]) to calculate the decay rates and energy shifts of SE from a resonance state.

In order to apply the non-Hermitian 2PT and extract the complex-valued energy correction due to SE, we rewrite (S1) as

$$\hat{H} = \hat{H}_0 + q\hat{H}_1 + q^2\hat{H}_2 \tag{S5}$$

with

$$\hat{H}_0 = \frac{\hat{p}^2}{2m} + V(x) + \hat{H}_R, \quad \hat{H}_1 = -\frac{1}{mc}\hat{A}\cdot\hat{p}, \quad \hat{H}_2 = \frac{1}{2mc^2}\hat{A}^2 . \tag{S6}$$

We will refer to $\hat{H}_0$ as the free Hamiltonian, and to $\hat{H}_1$ as the interaction Hamiltonian, since it exclusively describes the atom-photon interaction. $\hat{H}_2$ does not contribute to the transition amplitude between the initial and final states in 2PT, since the processes considered here involve a single photon. In higher order perturbation theories, however, this term will contribute.

We define eigenstates and eigenvalues (energies) of the free Hamiltonian from the respective eigenvalue problem

$$\hat{H}_0 |n\rangle = E^0_{n,\varphi(k)} |n\rangle . \tag{S7}$$

The eigenstates of the problem are written as a product of atomic and photonic parts, $|n\rangle = |\psi_n\rangle_A |\varphi(k)\rangle_R$, where the photonic state is either the photonic vacuum state $|vac\rangle_R$ for the initial atomic state, or a single photon state that can be written as $|\varphi(k)\rangle_R = \hat{a}^\dagger(k)|vac\rangle_R$ when the atom is in its final state. The eigenvalues of (S7) are the energies of the unperturbed system, and since we consider electronic bound states, these energies are real,

$$E^0_{n,\varphi(k)} = E^0_{A,n} + E^0_{ph,\varphi(k)}, \tag{S8}$$

where $E^0_{A,n}$ is the unperturbed energy of the n'th atomic level, and $E^0_{ph,\varphi(k)}$ is the unperturbed energy of the photon state $\varphi(k)$, which is zero if the photonic state is the vacuum state and $\hbar\omega_k$ otherwise, where $\omega_k$ is the frequency of a photon in a mode $k$. We remark that the framework developed here can be readily generalized for larger number of photons but in our 2PT analysis we do not consider multiphoton states.

We will now apply CS to the photonic coordinates. The detailed derivation of the CS of the photonic operators was done in [2], and here we will write the final expressions of the transformed operators (S2), (S3), (S5) and (S6) under the CS transformation rotation by an angle $\theta$:

$$\begin{aligned} \hat{a}(k) &\mapsto e^{i\theta/2}\hat{a}\left(ke^{i\theta}\right) \equiv \hat{a}_\theta(k), \\ \hat{a}^\dagger(k) &\mapsto e^{i\theta/2}\hat{a}^\dagger\left(ke^{i\theta}\right) \equiv \hat{b}^\dagger_\theta(k), \end{aligned} \tag{S9}$$

$$\hat{A}_\theta = \int_{-\infty}^{\infty} dk \sqrt{\frac{\hbar c^2}{8\pi\omega_k}}\hat{a}(k) + cc. = \hat{A}, \tag{S10}$$

$$\hat{B}_\theta = \int_{-\infty}^{\infty} dk \sqrt{\frac{\hbar c^2}{8\pi\omega_k}}\hat{a}^\dagger(k) + cc. = \hat{A}^\dagger, \tag{S11}$$

$$\hat{H}_{R,\theta} = e^{-i\theta}\hbar \int_{-\infty}^{\infty} dk\omega_k \hat{a}^\dagger(k)\hat{a}(k) = e^{-i\theta}\hat{H}_R, \tag{S12}$$

$$\hat{H}_{0,\theta} = \frac{\hat{p}^2}{2m} + V(x) + e^{-i\theta}\hat{H}_R, \tag{S13}$$

$$\hat{H}_{1,\theta} = -\frac{1}{mc}\hat{A}\cdot\hat{p}, \tag{S14}$$

$$\hat{H}_\theta = \hat{H}_{0,\theta} + q\hat{H}_{1,\theta} + q^2\hat{H}_{2,\theta}. \tag{S15}$$

Consequently, the energy of the photon will also be affected,

$$E^0_{ph,\varphi(k),\theta} = \begin{cases} 0, & \varphi(k) = vac \\ \hbar\omega_k e^{-i\theta}, & \varphi(k) = 1\ photon \end{cases} \tag{S16}$$

and correspondingly the CS unperturbed energy will be $E^0_{n,\varphi(k),\theta} = E^0_{A,n} + E^0_{ph,\varphi(k),\theta}$ and the complex-scaled eigenvector of the system will be $|n^\theta\rangle = |\psi^0_n\rangle_A|\varphi(k)\rangle_R$. Note that since the free radiation Hamiltonian $\hat{H}_R$ does not depend on the CS parameter $\theta$ (the dependence on $\theta$ is through a multiplicative factor $e^{-i\theta}$, however the constituting field raising and lowering operators are independent of $\theta$ as shown in (S12)), for the photonic basis set we use $|\varphi(k)\rangle_R = \hat{a}^\dagger(k)|vac\rangle_R$. We then expand (S15) with the electronic charge $q$ acting as a small parameter. The resulting 2PT correction to the energy of the $n$'th atomic level is given by

$$\begin{aligned} \delta E_n &\equiv \sum_{n'} \frac{\left|\left\langle n'^\theta \middle| q\hat{H}_{1,\theta} \middle| n^\theta \right\rangle\right|^2}{E^0_{n,0,\theta} - E^0_{n',1(k),\theta}} = \frac{\hbar q^2}{8\pi m^2}\sum_{n'}\int_{-\infty}^{\infty}\frac{dk}{\omega_k}\frac{|p_{nn'}|^2}{\hbar\omega_{nn'} - \hbar\omega_k e^{-i\theta}} \\ &= \frac{q^2}{8\pi m^2}\sum_{n'}|p_{nn'}|^2\int_{-\infty}^{\infty}\frac{dk}{\omega_k}\left[\frac{\omega_{nn'} - \omega_k\cos\theta}{(\omega_{nn'} - \omega_k\cos\theta)^2 + \omega_k^2\sin^2\theta} - i\frac{\omega_k\sin\theta}{(\omega_{nn'} - \omega_k\cos\theta)^2 + \omega_k^2\sin^2\theta}\right], \end{aligned} \tag{S17}$$

with the definition $\omega_{nn'} \equiv (E^0_{A,n} - E^0_{A,n'})\hbar^{-1}$ and the matrix element $p_{nn'} \equiv \langle\psi_{n'}|\hat{p}|\psi_n\rangle$.

Since the 2PT correction $\delta E_n = Re(\delta E_n) + i\,Im(\delta E_n)$ is a physical observable, it must be $\theta$ -independent, so without loss of generality we may consider $\theta \to 0^+$. Note that the correction (S17) is the sum of contributions from an infinite number of states, including the continuum solutions of the atomic part of the free Hamiltonian that we write for simplicity of notation as a sum rather than an integral. Then,

$$\mathrm{Re}(\delta E_n) = \frac{q^2}{4\pi m^2 c}\sum_{n'}|p_{nn'}|^2\, PV\left\{\int_0^\infty\frac{d\omega_k}{\omega_k}\frac{1}{\omega_{nn'} - \omega_k}\right\}, \tag{S18}$$

with $PV\{\,\}$ denoting Cauchy's principle integral and

$$\mathrm{Im}(\delta E_n) = -\frac{q^2}{4\pi m^2 c}\sum_{n'}|p_{nn'}|^2\int_0^\infty\frac{d\omega_k}{\omega_k}\pi\delta(\omega_{nn'} - \omega_k) = -\frac{q^2}{4m^2 c}\sum_{n'<n}\frac{|p_{nn'}|^2}{\omega_{nn'}}. \tag{S19}$$

The momentum operator matrix element $p_{nn'}$ can be re-written in terms of the dipole moment matrix element given by $d_{nn'} \equiv q\int_{-\infty}^{\infty}\psi^0_n(x)\,x\,\psi^0_{n'}(x)dx$

$$p_{nn'} = -i\frac{m}{\hbar}\left\langle\psi^0_n\middle|\left[\hat{x},\hat{H}_0\right]\middle|\psi^0_{n'}\right\rangle = i\frac{m}{q}\omega_{nn'}d_{nn'}. \tag{S20}$$

The SE decay rate is therefore defined as

$$\gamma_n^0 \equiv -2\mathrm{Im}\left(\delta E_n\right) = \sum_{n'<n} \frac{\omega_{nn'}}{2c} \left|d_{nn'}\right|^2 . \tag{S21}$$

This expression is the known result for a 1D problem SE decay rate [2], coinciding with the result of Fermi's golden rule. The energy shift (S18), however, is IR-divergent. To address that issue, we will first prove the following lemma.

*Lemma 1:* The sum

$$C_n = \sum_{n'=0}^{\infty} \frac{\left\langle \psi_n^0 \middle| \hat{p} \middle| \psi_{n'}^0 \right\rangle \left\langle \psi_{n'}^0 \middle| \hat{p} \middle| \psi_n^0 \right\rangle}{E_{A,n}^0 - E_{A,n'}^0} \tag{S22}$$

is a real quantity that does not depend on $n$.

*Proof:* We re-write the numerator of (S22)

$$\begin{aligned} \left\langle \psi_n^0 \middle| \hat{p} \middle| \psi_{n'}^0 \right\rangle \left\langle \psi_{n'}^0 \middle| \hat{p} \middle| \psi_n^0 \right\rangle &= -\frac{i}{2}\frac{m}{\hbar}\Big( \left(E_{A,n}^0 - E_{A,n'}^0\right) \left\langle \psi_n^0 \middle| \hat{p} \middle| \psi_{n'}^0 \right\rangle \left\langle \psi_{n'}^0 \middle| \hat{x} \middle| \psi_n^0 \right\rangle \\ &+ \left(E_{A,n'}^0 - E_{A,n}^0\right) \left\langle \psi_n^0 \middle| \hat{x} \middle| \psi_{n'}^0 \right\rangle \left\langle \psi_{n'}^0 \middle| \hat{p} \middle| \psi_n^0 \right\rangle \Big) \\ &= -i\frac{m}{2\hbar}\left(E_{A,n}^0 - E_{A,n'}^0\right)\Big( \left\langle \psi_n^0 \middle| \hat{p} \middle| \psi_{n'}^0 \right\rangle \left\langle \psi_{n'}^0 \middle| \hat{x} \middle| \psi_n^0 \right\rangle - \left\langle \psi_n^0 \middle| \hat{x} \middle| \psi_{n'}^0 \right\rangle \left\langle \psi_{n'}^0 \middle| \hat{p} \middle| \psi_n^0 \right\rangle \Big) . \end{aligned} \tag{S23}$$

Plugging this in (S22) and using the completeness property and the canonical commutation relation $[\hat{x}, \hat{p}] = i\hbar$,

$$\begin{aligned} C_n &= \frac{m}{2i\hbar} \sum_{n'} \Big( \left\langle \psi_n^0 \middle| \hat{p} \middle| \psi_{n'}^0 \right\rangle \left\langle \psi_{n'}^0 \middle| \hat{x} \middle| \psi_n^0 \right\rangle - \left\langle \psi_n^0 \middle| \hat{x} \middle| \psi_{n'}^0 \right\rangle \left\langle \psi_{n'}^0 \middle| \hat{p} \middle| \psi_n^0 \right\rangle \Big) \\ &= \frac{m}{2i\hbar} \left( \left\langle \psi_n^0 \middle| \hat{p} \sum_{n'} \middle| \psi_{n'}^0 \right\rangle \left\langle \psi_{n'}^0 \middle| \hat{x} \middle| \psi_n^0 \right\rangle - \left\langle \psi_n^0 \middle| \hat{x} \sum_{n'} \middle| \psi_{n'}^0 \right\rangle \left\langle \psi_{n'}^0 \middle| \hat{p} \middle| \psi_n^0 \right\rangle \right) \\ &= \frac{m}{2i\hbar} \left\langle \psi_n^0 \middle| \hat{p}\hat{x} - \hat{x}\hat{p} \middle| \psi_n^0 \right\rangle = -\frac{m}{2} \end{aligned} \tag{S24}$$

which is a real number that does not depend on $n$. ■ We emphasize that this proof is valid only when the complete basis set of the atomic states is used, in order to apply the completeness property. In other words, all atomic transitions must be considered for this proof to hold.

We will now prove that the IR-divergence of (S18) is physically irrelevant, since the divergent contribution to the energy shift is equal for all atomic levels. Consider the energy-shift difference between any two atomic levels $n_1$ and $n_2$,

$$Z_{n_1 n_2} = \mathrm{Re}\left(\delta E_{n_2}\right) - \mathrm{Re}\left(\delta E_{n_1}\right) = \frac{1}{4\pi c}\left( \sum_{n_2} \left|d_{nn_2}\right|^2 \omega_{nn_2}^2 PV\left\{ \int_0^\infty \frac{d\omega_k}{\omega_k} \frac{1}{\omega_{nn_2} - \omega_k} \right\} - \sum_{n_1} \left|d_{nn_1}\right|^2 \omega_{nn_1}^2 PV\left\{ \int_0^\infty \frac{d\omega_k}{\omega_k} \frac{1}{\omega_{nn_1} - \omega_k} \right\} \right) . \tag{S25}$$

The IR-divergent part emerges for $\omega_k \ll \omega_{nn'}$ (for any $n'$). In this limit (S25) becomes

$$Z_{n_1 n_2} \approx \frac{1}{4\pi c}\left(\sum_{n_2}\left|d_{nn_2}\right|^2 \omega_{nn_2} - \sum_{n_1}\left|d_{nn_1}\right|^2 \omega_{nn_1}\right) PV\left\{\int_0^\infty \frac{d\omega_k}{\omega_k}\right\}. \quad \text{(S26)}$$

However, from the Lemma 1 c.f. (S20), equation (S26) vanishes. Therefore, the IR-divergence has no physical meaning and we can define the measurable energy shift as the expression (S18) minus the IR-divergent part of the Cauchy's principle integral,

$$\Delta_n \equiv \frac{1}{4\pi c}\sum_{n'}\left|d_{nn'}\right|^2 \omega_{nn'}^2 PV\left\{\int_0^\infty \frac{d\omega_k}{\omega_k}\frac{1}{\omega_{nn'}-\omega_k} - \int_0^\infty \frac{d\omega_k}{\omega_{nn'}\omega_k}\right\} = \frac{1}{4\pi c}\sum_{n'}\left|d_{nn'}\right|^2 \omega_{nn'} PV\left\{\int_0^\infty \frac{d\omega_k}{\omega_{nn'}-\omega_k}\right\}. \quad \text{(S27)}$$

The log-divergence of this expression has been at the center of much debate in the early days of quantum-mechanical theory, and it is commonly alleviated by applying Bethe's 1947 regularization, namely – truncating the integral with a cutoff at the Compton energy $mc^2$, above which the electron must be treated relativistically [1]. Replacing the upper integration limit $\infty$ with $\Omega = mc^2/\hbar$ finally yields the measurable energy shift

$$\Delta_n = \frac{1}{4\pi c}\sum_{n'}\left|d_{nn'}\right|^2 \omega_{nn'} \ln\left(\frac{\hbar\left|\omega_{nn'}\right|}{mc^2}\right). \quad \text{(S28)}$$

The choice of cutoff frequency is also physically inconsequential for calculating the total SE decay rate from the atomic level $n$, since it shifts $\Delta_n$ by the same amount independently of $n$. However, when considering the decay rate from the initial state $n$ to any specific final state $n'$ this cutoff is a measurable quantity.

We remark that while the SE decay rate (S21) depends only on lower-energy states (in fact, the SE rate of level $n$ is the sum of the partial decay rates into all atomic levels $n' < n$), the measurable energy shift (S28) is also affected by higher-energy states, as well as by the electronic continuum. As discussed in the main text and will be shown in the next supplementary section, this is not the case for SE from ERS – where SE from every state is affected by every state and the entire continuum.

## C. Spontaneous emission from electronic resonance states in the NHQM formalism

We will now proceed to calculate the SE decay rates and energy shifts from ERS. The generalization is relatively straightforward, and entails replacing the real-valued atomic energies with complex-valued energies, which are the eigenvalues of the non-Hermitian atomic Hamiltonian describing metastable (resonant) electronic states.

The rotated free non-Hermitian Hamiltonian $\hat{H}_{0,\theta}$ where the electronic coordinates are rotated by an angle $\theta$ and the photonic coordinates by $2\theta$ (as explained in Supplementary Section A) is

$$\hat{H}_{0,\theta} = \frac{\hat{p}^2 e^{-2i\theta}}{2m} + V\left(xe^{i\theta}\right) + e^{-2i\theta}\hat{H}_R \quad \text{(S29)}$$

and its corresponding right- and left-eigenvectors are defined, respectively,

$$\hat{H}_{0,\theta}|n^\theta) = E^0_{n,\theta,\varphi(k)}|n^\theta), \quad \text{(S30)}$$

$$(n^\theta|\hat{H}_{0,\theta} = (n^\theta|E^0_{n,\theta,\varphi(k)}, \tag{S31}$$

with $|n^\theta) = |\psi_n^{0,\theta})_A|\varphi(k)\rangle_R$, using the vector notation satisfying the c-product for the electronic states, namely $(\psi_1|\psi_2) = \int_{\mathbb{R}^1}\psi_1(x)\psi_2(x)dx$, rather than the standard vector product [1]. For the photonic state, we may use the unscaled basis set $|\varphi(k)\rangle = \hat{a}^\dagger(k)|vac\rangle$ (as in Supplementary Section B) since the radiation Hamiltonian $\hat{H}_R$ in (S29) does not depend on the CS angle $\theta$ (see eq. (S12)).

We write the state vectors $|n^\theta) = |\psi_n^{0,\theta})_A|\varphi(k)\rangle_R$ and the complex atomic energy of the $n$'th level $E_n^A \equiv \hbar\omega_n - i\Gamma_n/2$, with $\Gamma_n$ the ERS linewidth. An illustration of the rotated complex spectrum of the unperturbed Hamiltonian (S29), including both the electronic and photonic continua, is shown in Fig. S1.

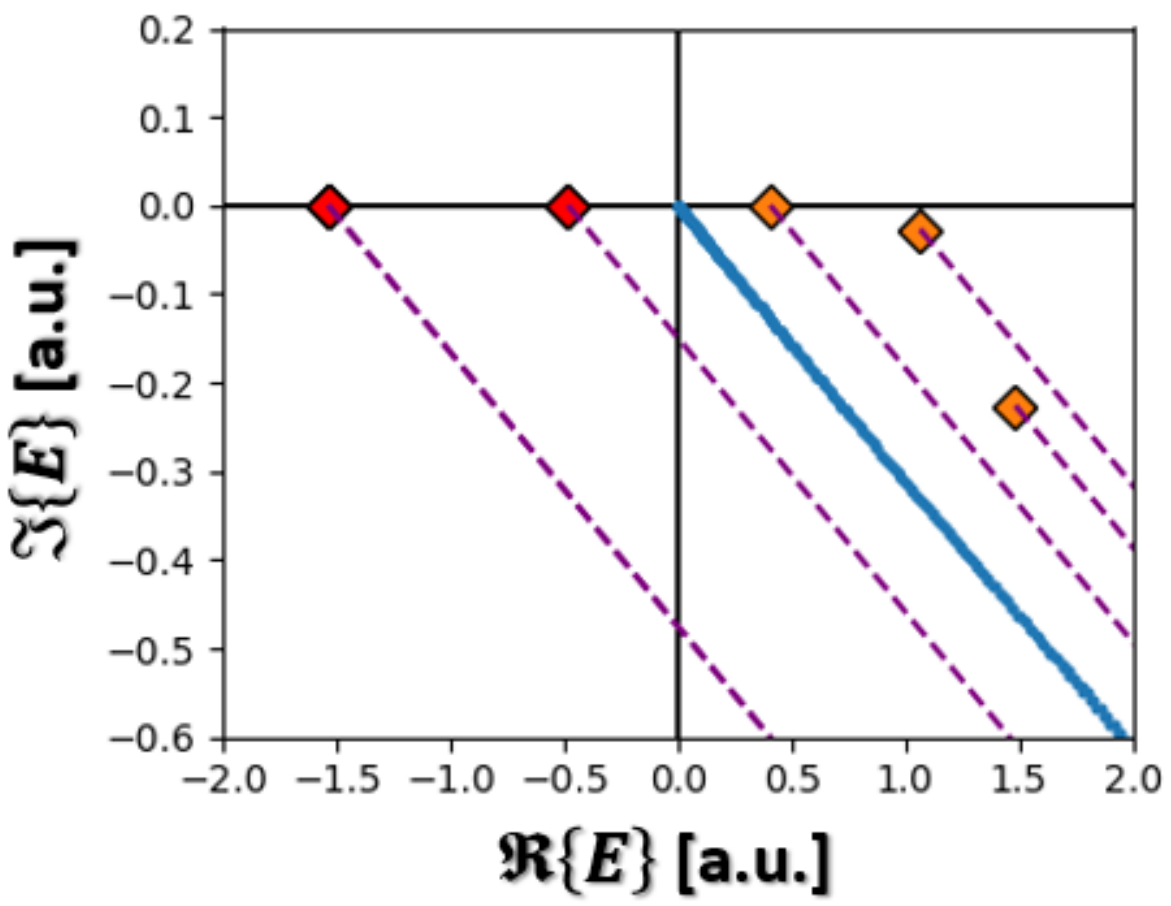

Fig. S1. Electronic and photonic complex spectrum of the rotated unperturbed Hamiltonian $\hat{H}_{0,\theta}$ for $\theta = 0.15[rad]$ and $V(x) = (0.5x^2 - 2.1)e^{-0.1x^2}$ (in a.u. – atomic units). Red (orange) diamonds denote the bound (resonance) states. Blue line denotes the rotated electronic continuum and purple dashed lines denote the photonic rotated continua.

We repeat what was done in Supplementary Section B by employing 2PT and obtain

$$\delta E_n^{res} = \frac{\hbar q^2}{8\pi m^2}\sum_{n'}\int_{-\infty}^{\infty}\frac{dk}{\omega_k}\frac{p_{nn'}p_{n'n}}{\hbar\omega_{nn'} - \hbar\omega_k e^{-2i\theta} - i\left(\Gamma_n - \Gamma_{n'}\right)/2}. \tag{S32}$$

Note that the momentum matrix element is still $\theta$ -independent,

$$p_{nn'} = \left(\psi_n^{0,\theta}\middle|\hat{\mathbf{p}}e^{-i\theta}\middle|\psi_{n'}^{0,\theta}\right) = -i\frac{m}{q\hbar}\left(\psi_n^{0,\theta}\middle|\left[\hat{\mathbf{d}}e^{i\theta}, \hat{H}_{0,\theta}\right]\middle|\psi_{n'}^{0,\theta}\right) = i\frac{m}{q}f_{nn'}\mathbf{d}_{nn'} \tag{S33}$$

where we have introduced the complex frequency,

$$f_{nn'} \equiv \omega_{nn'} - i\left(\Gamma_n - \Gamma_{n'}\right)(2\hbar)^{-1}. \tag{S34}$$

Due to the CS transformation of the wavefunction $\psi_n^0 \mapsto e^{i\theta/2}\psi_n^0\left(xe^{i\theta}\right)$ [1], the dipole matrix element from equation (S33) is given by

$$\mathbf{d}_{nn'} = \left( \psi_n^{0,\theta} \middle| \hat{\mathbf{d}} e^{i\theta} \middle| \psi_{n'}^{0,\theta} \right) = q \int_{-\infty}^{\infty} dx e^{i\theta} \psi_n^0 \left( x e^{i\theta} \right) e^{i\theta/2} x e^{i\theta/2} \psi_{n'}^0 \left( x e^{i\theta} \right) \tag{S35}$$

and by defining the atomic rotated coordinate $\xi \equiv x e^{i\theta}$ we integrate along the rotated real axis $l_\theta$

$$\mathbf{d}_{nn'} = q \int_{l_\theta} d\xi \psi_n^0 (\xi) \xi \psi_{n'}^0 (\xi) . \tag{S36}$$

Substituting $\omega_k = c|k|$ and defining the photonic rotated coordinate $\zeta \equiv \omega_k e^{-2i\theta}$ we write (S32) as

$$\delta E_n^{res} = \frac{1}{4\pi c} \sum_{n'} \mathbf{d}_{nn'}^2 \int_{l_{2\theta}} \frac{d\zeta}{\zeta} \frac{f_{nn'}^2}{f_{nn'} - \zeta} , \tag{S37}$$

where the integration is performed along $l_{2\theta}$, the positive real semi-axis rotated clockwise by $2\theta$.

One can immediately see the resemblance to the expression for the 2PT correction in the EBS case. However, equation (S37) cannot be decomposed into a real Cauchy principle integral and an imaginary number due to the rotation; in the present case, we are not allowed to take the limit $\theta \to 0^+$ because the resonance states lie in the fourth quadrant of the complex plane. Therefore, when calculating the 2PT energy correction of the $n$ ERS, the CS rotation angle is subjected to the condition $\theta > \frac{1}{2}\arctan(\Gamma_n/2\hbar\omega_n)$.

As in the EBS case, the real part equation (S37) is also IR-divergent. The IR-divergent part can be discarded in the same manner as before by taking the limit $|\zeta| \ll |f_{nn'}|$ for any final state $n'$ in (S37) and applying Lemma 1 (which is still valid even when considering the non-Hermitian version as can be easily affirmed) with the definition (S34). Analogously to (S27), we define the non-divergent part of the 2PT correction by subtracting the $\theta$-dependent quantity

$$\Xi_n(\theta) = \frac{1}{4\pi c} \sum_{n'} \mathbf{d}_{nn'}^2 \int_{l_{2\theta}} \frac{f_{nn'}}{\zeta} d\zeta \tag{S38}$$

from (S37). Note that unlike the Hermitian case, here the IR-divergent part is a complex number. We obtain the non-IR-divergent correction

$$\delta \tilde{E}_n^{res}(\theta) \equiv -\lim_{|s_\infty|\to\infty} \sum_{n'} \frac{d_{nn'}^2 f_{nn'}}{4\pi c} \left[ Log\left( f_{nn'} - \zeta \right) \Big|_0^{s_\infty} \right] = -\lim_{|s_\infty|\to\infty} \sum_{n'} \frac{d_{nn'}^2 f_{nn'}}{4\pi c} \left( \log|s_\infty| - 2i\theta - \log|f_{nn'}| - i \arg\left(-f_{nn'}\right) \right) \tag{S39}$$

using the complex logarithm defined as $Log(z) \equiv \log|z| + i\arg(z)$ for $z \in \mathbb{C}$. The ambiguity in the phase is alleviated by the fact that the integrand is analytic along the $l_{2\theta}$ line and we may vary $s_\infty$ continuously. We also require that when we approach the bound-state case, the energy shift and linewidth converge to the expressions (S28) and (S21) respectively. Note that expression (S39) depends on the CS parameter $\theta$; however, this dependence is an artifact introduced by the subtraction we have performed to regularize the IR divergence, as we will now show.

*Lemma 2*: The origin of the $\theta$ dependence of (S39) stems solely from the IR regularization. In other words, for any two clockwise rotation angles $2\theta_1$ and $2\theta_2$ about the complex origin, equation (S38) satisfies the relation $\Xi_n(\theta_1)-\Xi_n(\theta_2)=\frac{i}{4\pi c}\sum_{n'}\mathbf{d}_{nn'}^2 f_{nn'}(2\theta_2-2\theta_1)$, which is equal to $\left[-\left(\delta\tilde{E}_n^{res}(\theta_1)-\delta\tilde{E}_n^{res}(\theta_2)\right)\right]$.

*Proof:* Consider a single term from the sum (S38),

$$\Xi_{nn'}(\theta)=\frac{1}{4\pi c}\mathbf{d}_{nn'}^2\int_{l_{2\theta}}\frac{f_{nn'}}{\zeta}d\zeta\ . \tag{S40}$$

We wish to show that $\Xi_{nn'}(\theta_1)-\Xi_{nn'}(\theta_2)=\frac{i}{4\pi c}\mathbf{d}_{nn'}^2 f_{nn'}(2\theta_2-2\theta_1)$ for any clockwise rotation angle $2\theta_1,2\theta_2$ about the complex origin. To that end, we construct a contour consisting of two semi-infinite axes extending from the complex origin rotated counterclockwise by $2\theta_1$ and $2\theta_2$, connected by an arch at $|\zeta|\to\infty$ as illustrated in Fig. S2. In this proof we will avoid the divergence at $\zeta\to 0$ by introducing an auxiliary parameter $\varepsilon\to 0^+$ such that $\zeta\to(\zeta^2+\varepsilon^2)^{1/2}$, thereby replacing the divergence at origin with two singularities at $\pm i\varepsilon$ which lay outside this contour. Since there are no poles inside this contour, by using Cauchy's theorem we have (without loss of generality, we assume $\theta_2>\theta_1$)

$$0=\Xi_{nn'}(\theta_2)-\frac{1}{4\pi c}\mathbf{d}_{nn'}^2\int_R\frac{f_{nn'}}{\zeta}d\zeta-\Xi_{nn'}(\theta_1)\,, \tag{S41}$$

where $R$ is the arch extending between the two rotated semi-infinite axes at $|\zeta|\to\infty$. We denote the intersection points of the arch with the semiaxis rotated by $2\theta_j$ by $s_{\infty,j}$, where $j=\{1,2\}$ and $|s_{\infty,1}|=|s_{\infty,2}|$ is taken to be infinitely large. Then,

$$\frac{1}{4\pi c}\mathbf{d}_{nn'}^2\int_R\frac{f_{nn'}}{\zeta}d\zeta=\frac{1}{4\pi c}\mathbf{d}_{nn'}^2\,Log(f_{nn'}-\zeta)\Big|_{s_{\infty,1}}^{s_{\infty,2}}=\frac{1}{4\pi c}\mathbf{d}_{nn'}^2\left(\log|s_{\infty,2}|-\log|s_{\infty,1}|-2i\theta_2+2i\theta_1\right)=\frac{i(2\theta_1-2\theta_2)}{4\pi c}\mathbf{d}_{nn'}^2\,. \tag{S42}$$

Plugging this in (S41) and summing over all $n'$ completes the proof. ■

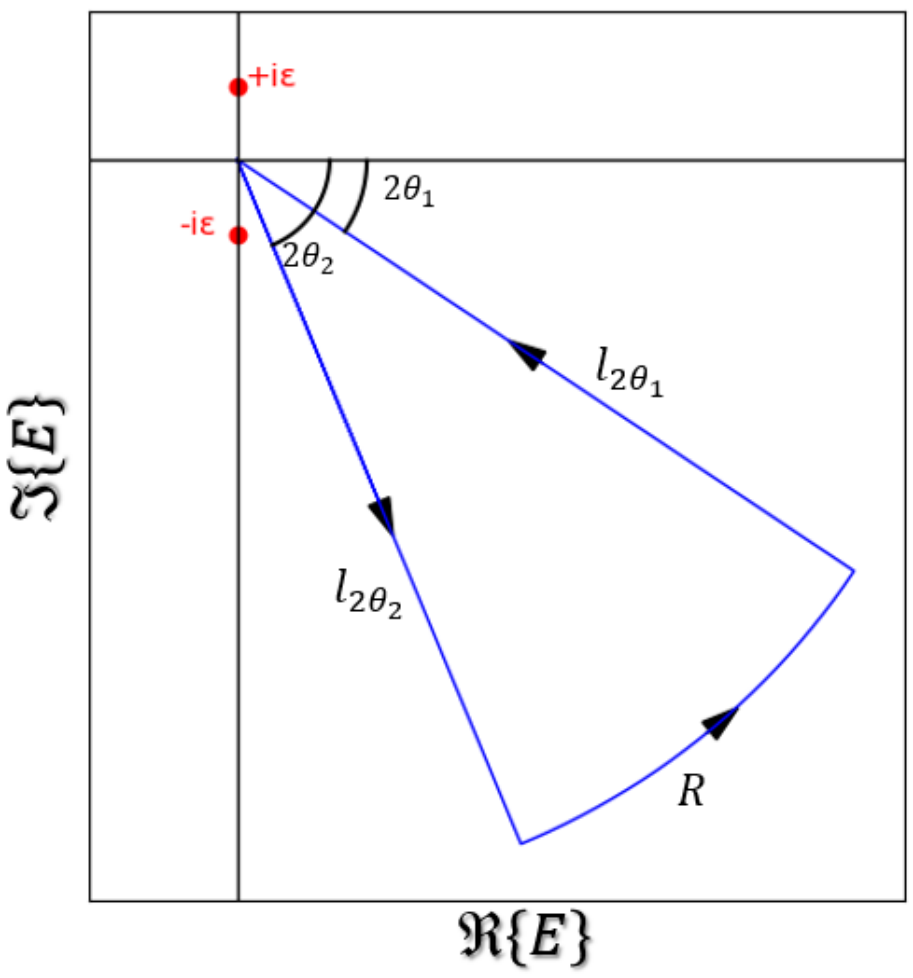

Fig. S2. Contour integration (equation (S41)) along two semi-infinite axes rotated clockwise by $2\theta_1$ and $2\theta_2$ about the complex origin and connected by an arch at infinity.

To address the logarithmic divergence in (S39), we apply again a frequency cutoff commensurate with the Compton energy, namely $|s_\infty| = mc^2/\hbar$ in analogy to the Bethe treatment applied to the Hermitian case. As in the Hermitian case of Supplementary Section B, the cutoff frequency has no consequence when considering the total decay rate; however, it is meaningful when calculating the individual contributions to the total SE decay rate.

Finally, after setting a cutoff frequency and ignoring the physically meaningless dependence of (S39) on $\theta$, we define the measurable 2PT correction for ERS,

$$\delta\bar{E}_n^{res} \equiv \sum_{n'} \frac{Z_{nn'}}{4\pi c}\left(\log\frac{|f_{nn'}|\hbar}{mc^2} + i\arg(-f_{nn'})\right). \tag{S43}$$

For brevity, we denote $Z_{nn'} = d_{nn'}^2 f_{nn'}$. Since we wish to make a clear distinction between the discrete states (ERS and EBS) and the continuum, we shall separate the sum in (S43) into a summation over all $N(\theta)$ discrete states (the number of which depends on $\theta$), and a summation over the continuum states which we will write as an integral in the limit of an infinite number of states,

$$\delta\bar{E}_n^{res} \equiv \sum_{n'}^{N(\theta)} \frac{Z_{nn'}}{4\pi c}\left(\log\frac{|f_{nn'}|\hbar}{mc^2} + i\arg(-f_{nn'})\right) + \int_{l_{2\theta}} d\mathcal{E}\,\frac{Z_n(\mathcal{E})}{4\pi c}\left(\log\frac{|f_n(\mathcal{E})|\hbar}{mc^2} + i\arg(-f_n(\mathcal{E}))\right). \tag{S44}$$

Note that the integral depends on $N(\theta)$ since $\delta\bar{E}_n^{res}$ must be independent of $\theta$; whenever another resonance pole emerges from the continuum, the value of the integral changes so that in total $\delta\bar{E}_n^{res}$ remains constant. In the above we made the analogous definitions $Z_n(\mathrm{E}) = d_n^2(\mathrm{E}) f_n(\mathrm{E})$, $f_n(\mathrm{E}) = \tilde{\omega}_n(\mathrm{E}) - i\tilde{\Gamma}_n(\mathrm{E})/2\hbar$, $\tilde{\omega}_n(\mathrm{E}) = \omega_n - Re(\mathrm{E})/\hbar$ and $\tilde{\Gamma}_n(\mathrm{E}) = \Gamma_n + 2\,Im(\mathrm{E})$, where $d_n(\mathrm{E}) = q\int_{l_{2\theta}} \psi_n^\theta(z)\, z\psi^\theta(\mathrm{E}, z)dz$ is the transition dipole moment between a discrete state $n$ and the continuum "state" with energy $\mathcal{E}$. The ERS energy shift due to SE and the SE decay rate from the $n$'th atomic level are respectively

$$\Delta_n^{res} \equiv \frac{1}{4\pi c}\sum_{n'}^{N(\theta)}\left[\mathrm{Re}\left(Z_{nn'}\right)\log\frac{\left|f_{nn'}\right|\hbar}{mc^2} - \mathrm{Im}\left(Z_{nn'}\right)\arg\left(-f_{nn'}\right)\right]$$
$$+\frac{1}{4\pi c}\int_{l_{2\theta}} d\mathcal{E}\left[\mathrm{Re}\left(Z_n(\mathcal{E})\right)\log\frac{\left|f_n(\mathcal{E})\right|\hbar}{mc^2} - \mathrm{Im}\left(Z_n(\mathcal{E})\right)\arg\left(-f_n(\mathcal{E})\right)\right], \tag{S45}$$

$$\gamma_n = -\frac{1}{2\pi c}\sum_{n'}^{N(\theta)}\left[\mathrm{Im}\left(Z_{nn'}\right)\log\frac{\left|f_{nn'}\right|\hbar}{mc^2} + \mathrm{Re}\left(Z_{nn'}\right)\arg\left(-f_{nn'}\right)\right]$$
$$-\frac{1}{2\pi c}\int_{l_{2\theta}} d\mathcal{E}\left[\mathrm{Im}\left(Z_n(\mathcal{E})\right)\log\frac{\left|f_n(\mathcal{E})\right|\hbar}{mc^2} + \mathrm{Re}\left(Z_n(\mathcal{E})\right)\arg\left(-f_n(\mathcal{E})\right)\right]. \tag{S46}$$

In atomic units one retrieves equations (1)-(3) in the main text from (S46). In order to reproduce equations (S28) and (S21) in the case $f_{nn'} \to \omega_{nn'}$, namely in the Hermitian limit, we choose the branch of the $\arg(z)$ function for a complex quantity $z \in \mathbb{C}$ such that $\arg(-z) = \arg(z) - \pi$. In the Hermitian limit, the transition dipole moment matrix elements are real and therefore $\mathrm{Im}(Z_{nn'}) \to 0$ and $\arg(-f_{nn'}) \to -\pi\Theta(n - n')$ with $\Theta(x)$ the Heaviside distribution, thereby reproducing (S21). We see again that in the Hermitian case, the total decay rate from a bound state $n$ is contributed to only by the partial decay channels corresponding to transitions into lower energy bound states $n' < n$ whereas the energy shift is affected by all discrete and continuum states. However, when $f_{nn'}, d_{nn'} \in \mathbb{C}$ both the SE decay rate and energy shift of an atomic level $n$ depend on all discrete states and on the entire continuum.

## D. Implementation example – semiconductor quantum well emitting at mid-IR frequency

We show here that the potential function discussed in the main text (similar to a potential that was used to describe resonant tunnelling in a diode [29]) provides a reasonable approximation of realizable quantum emitters based on epitaxially grown semiconductor structures. Specifically, we consider intraband spontaneous emission from a metastable state to the ground state within a semiconductor quantum well (QW) with finite tunneling barriers (Fig S3(a)). Such a light-emitting system can not only be readily designed and fabricated to experimentally verify our model predictions– but also serves as a key component in quantum cascade lasers, which rely on light emission from metastable states [8]. The potential function (in atomic units)

$$V(x) = \frac{1}{100}\left(0.5\left(\frac{x}{\sqrt{200}}\right)^2 - 1.2\right)\exp\left(-0.12\left(\frac{x}{\sqrt{200}}\right)^2\right) \tag{S47}$$

corresponds to a QW approximately $0.327[eV]$ deep and $1.95[nm]$ wide, with barriers of height of approximately $0.312[eV]$ and width of $2.4[nm]$ as illustrated in Fig S3(a). Such parameters are attainable in AlGaN/GaN based QWs with tunnelling barriers [9]. Assuming an effective electron mass of $m^* \approx 0.25m_0$ [10], This potential well supports a single bound state with energy $E_0 \approx -0.131[eV]$ with respect to the threshold energy.

We focus here on the spontaneous emission from the first resonance state with energy $E_1 \approx 0.195[eV]$ above the threshold energy into the ground state, which upon decaying to the ground state emits a photon corresponding to a mid-IR wavelength of $\lambda \approx 3.8[\mu m]$. Applying CS to the electronic Hamiltonian with the potential (S47) and calculating the decay rate (S46), we show in Fig. S3(b,c) that the decay rate calculated from the ground state alone deviates by approximately 2.13% from the exact value obtained from integration of the entire continuum, while calculations of the contributions of eight discrete states converges to within 0.06% of the exact value.

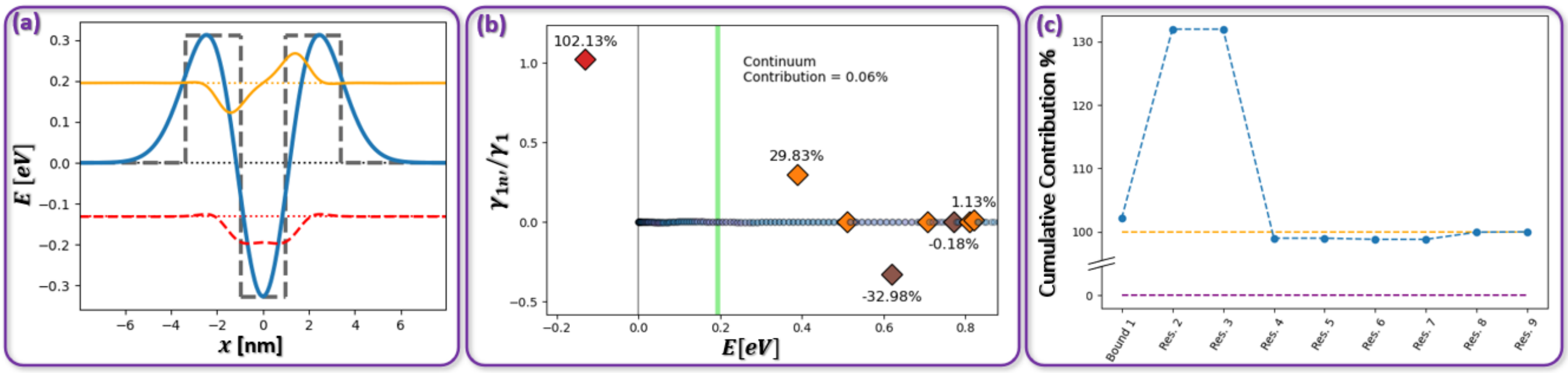

Fig. S3. (a) Potential describing a semiconductor quantum well (S47) supporting a single ground state (dashed red line). Orange solid line marks the first resonance state. The wavefunctions are calculated at a complex-scaling angle $\theta = 0.62[rad]$. Grey dashed line outlines the corresponding approximation of a rectangular AlGaN/GaN QW potential (b) Decomposition of the total radiative decay rate of the first resonance state $\gamma_1$ into discrete states and the continuum. Approximately 99.94% of the SE decay rate can be calculated from only the first eight discrete states. (c) cumulative contribution of the SE decay into the discrete states to the total SE decay rate of the first resonance state, $\gamma_{1n'}$.